# Novel Method for More Efficient Optimization of Knowledge-Based Planning: Specific Voxels of each Structure Influenced by Dominant Beamlets (SVSIDB)

Ali Yousefi[1*], Saeedeh Ketabi[1], Iraj Abedi[2]

[1] Department of Management- Operations Research, University of Isfahan, Isfahan, Iran
[2] Department of Medical Physics, School of Medicine, Isfahan University of Medical Sciences, Isfahan, Iran
[*] Corresponding Author- email address: yousefi@ase.ui.ac.ir

**Abstract**
Background: There is a huge problem and time-consuming computation to optimize the intensity-modulated radiotherapy (IMRT) treatment plan. Extracting the optimized plan from the predicted three dimensions does distribution ($3D^3$) so-called optimizing the knowledge-based planning (KBP) is also involved in this challenge. Some algorithms and methods have been presented for clustering and down-sampling the voxels to make the problem smaller, in recent years.
Purpose: In the current research, a novel down-sampling method is presented for optimizing the knowledge-based planning more efficiently.
Methods: The concept of Specific Voxels of each Structure Influenced by the Dominant Beamlet (SVSIDB) and corresponding down-sampling algorithm are proposed under title of SMP-2. The algorithm has been run on the data of 30 patients with head and neck tumors randomly selected from the Open-KBP dataset. For each patient, there are 19 sets of dose prediction data in this dataset. Therefore, a total of 570 KBP-optimizing problems have been solved by applying the QuadLin model in the CVX framework. Resulted plans are evaluated and compared regarding two main fields which are the quality of the treatment plan as well as the computation efficiency. The former consists of two plan quality approaches including dose-volume histogram (DVH) points differences, and satisfied clinical criteria. Solve time is the evaluation criteria for the latter field i.e. computation efficiency.
Results: The results of the current study indicated a remarkable improvement in the computation efficiency compared to the results of previous research while at the same time outperforming the plan evaluation indicators. Accordingly, the proposed method, SMP-2, reduced the average solving time by 46% in comparison to the full-data QuadLin model. The results also show an up to 53% reduction in solve time along with up to 22% improvement in clinical criteria compared to the previous research.
Conclusions: Evaluation of the research results indicated that the *SVSIDB* has not only reduced the solve time but also improved the quality of the treatment plans. This is a remarkable achievement of the proposed model compared to the previous research and confirmed the significant effectiveness of the *SVSIDB* method which has the potential of even more improvement of the computation efficiency.

## I. Introduction

Optimizing the IMRT treatment plan is time-consuming due to its large number of voxels. Most of the studies on solving such problems have implemented random down-sampling methods to reduce the size of the problem [1]. As one of the pioneers in this field, Kufer et al. (2003) proposed a two-step process to get along with large-scale problems, the first step includes identifying and eliminating unnecessary constraints. The second step successively combines similar constraints in clusters, therefore the resulting optimization problem is a fine approximation of the primary problem. At the end, the optimal solution to the original problem was found with a local refinement method [2]. Prior studies in this field have investigated the resolution of the voxel grid, in both geometrically regular blocks and irregular—but geometrically connected—blocks by clustering based on the value of corresponding rows of voxels in the dose influence matrix [3,4]. These studies introduce techniques for the adaptive refinement of groups to make better the quality of the approximate solution. Ferris et al. (2006) proposed a multi-phase sampling to lessen solving time and at the same time maintain the solution quality [5]. Martin et al.'s (2007) study on a stochastic optimization of IMRT using voxel sampling introduced a different method in this field. they assign voxels to cluster regions, then applied a stochastic gradient descent that focuses on dose calculations and evaluations. The objective function relies on a small number of voxels randomly sampled from each region in each iteration [6]. As expected, the solution returned by this algorithm converges to the solution of the complete problem [7]. Furthermore, Lan et al. (2012) applied a geometric distance sorting method for voxel-selecting to be iteratively added to the constraints for quadratic optimization models that include DVH constraints [8]. Potrebko et al. (2017) provided a sampling method to improve the efficiency of large-scale problems. Their method applies Pareto-aware radiotherapy evolutionary treatment optimization to optimize the fluence patterns and beam angles [9]. Some studies presented different approaches to clustering the voxels applying meta heuristic algorithms [10,11].

---
[1] email address: yousefi@ase.ui.ac.ir

Inspired by the K-means algorithm, Mahnam et al. (2017) presented a model for clustering voxels based on similarity with neighboring voxels. In their proposed method, first, the number of clusters is determined, voxels are randomly assigned to clusters and the center of each cluster is defined as the average of observations. The distance of the voxel from its neighboring clusters is calculated because it is expected that there is a high degree of similarity between the neighbors. The average Euclidean distance of voxels from the center of the corresponding cluster was considered as a measure of clustering quality [1]. Moreover, integrating the K-means algorithm with metaheuristic techniques is led to efficient clustering methods [12]. Yang and Xing (2018) used isodose-feature preserving voxelization to establish clusters of voxels on the base of the treatment plans of the previous patients. This method down-sizes the number of voxels, consequently making smaller the number of constraints [13]. In the most recent research, Fountain et al. (2021) presented a dose-based technique for constraint generation to solve large-scale linear problems in IMRT. First, they used the special features of the IMRT problem, i.e. dose influence matrix, to cluster voxels based on how each part is affected by beams of unit intensity, and then used these clusters in a constraint generation algorithm. Their clustering approach is that all the voxels in which the dominant beamlet (the beamlet that delivers the highest dose to the voxel) is specific, are placed in one cluster and one voxel is considered as the representative of the cluster. The applicability of the proposed method was demonstrated using a retrospective dataset of 8 breast cancer patients with an organ at risk (OAR), one tumor volume, and two beams [14].

The main idea in Knowledge-Based Planning is to extract a model for predicting appropriate planning goals from a previously planned patient database. KBP uses a database of previously treated patients to create a model that matches the dose delivered to the patient's voxels with the final dose-volume histogram. In this regard, Momin et al. (2021) [15] provided a survey on KBP methods during the last decade which classified into two major categories consisted of traditional KBP methods [16-20] and deep-learning methods [21-28]. In recent years, with the development of deep learning techniques, it has become possible to predict the three-dimensional dose distribution ($3D^3$) of a new patient based on the treatment plans of similar recent patients. Several research has been done on the optimization of knowledge-based planning using different deep-learning algorithms and dose-mimicking models [29-36]. Most of them attempted to produce a treatment plan closer to the predicted dose i.e. they tried to pull the dose in each voxel to its predicted value. Babier et al. (2022) presented four optimization models (MaxAbs, MeanAbs, MaxRel, and MeanRel) with a dose-mimicking approach and applied the open-KBP dataset. They applied some plan evaluation approaches including DVH points differences and clinical criteria satisfaction and illustrated an increase of about 6% in the satisfaction of clinical indicators compared to the predicted dose with an average solve time of 343 seconds [37]. Yousefi et al. (2023) proposed a semi-automated mathematical dose-mimicking model to generate a Pareto optimal IMRT treatment plan using a combination of the $3D^3$ prediction and dose prescription values, titled QuadLin model. They implemented the QuadLin on the open-KBP dataset and improved the quality of generated treatment plans in both clinical criteria and computational efficiency with at least 13% better results than Babier et al. (2022) in clinical indicators, and an average solve time of 341 seconds [38].

In the current research, a novel method is presented for optimization of knowledge-based planning more efficiently by defining the concept of Specific Voxels of each Structure Influenced by the Dominant Beamlet (SVSIDB) and proposing the corresponding down-sampling algorithm. The method has been implemented on the data of 30 patients with head and neck tumors randomly selected from the Open-KBP dataset in order to properly compare the results with the previous research by Yousefi et al. (2023). For each patient, there are 19 sets of dose prediction data in this dataset, hence, a total of 570 KBP-optimizing problems have been solved by applying the QuadLin model in the CVX framework. The resulting plans titled by SMP-2 are evaluated and compared with the results of previous research, including down-sampled data (SMP-1) and full-data (MaxAbs, MeanAbs, MaxRel, MeanRel, and full-data QuadLin model), regarding two main fields which are the treatment quality as well as the computation efficiency. The former consists of two plan quality approaches including DVH points differences, and satisfied clinical criteria. Solve time is the evaluation criteria for the latter field i.e. computation efficiency. Our results indicate a strong significant improvement in both computation efficiency and plan evaluation indicators, compared to the results of previous research. More clearly, the proposed method, SMP-2, reduced the average solving time by 46% in comparison to the full-data QuadLin model by Yousefi et al. (2023), an up to 53% reduction in solve time along with an average 12% improvement in clinical criteria compared with the full-data models by Babier et al. (2022), and also more than 22% betterment of satisfying clinical criteria in comparison with down-sampled data SMP-1 by Fountain et al. (2021).

## II. Methods
### A. Down-Sampling algorithm

In this paper, a new algorithm for voxels down-sampling (clustering) is presented by defining the concept of *Specific Voxels of each Structure Influenced by Dominant Beamlet* (SVSIDB). We know that each voxel is influenced by several beamlets, and the beamlet that has the most influence on the voxel –based on the dose influence matrix– is called the dominant beamlet of that voxel. On the other hand, each beamlet is a dominant beamlet for a number of voxels; it means the beamlet has the greatest influence on that voxels compared to other beamlets. Without losing generality, the influence of beamlets on the voxels of each structure is investigated independently and separately from other structures. Accordingly, assuming that a specific beamlet is the dominant beamlet for a set of voxels in each of the structures; at this stage, some voxels of that set can be chosen as samples related to that beamlet and structure.

The main goal of the IMRT treatment planning problem is to determine the optimal intensity of beamlets to achieve the prescribed dose for the tumor as well as protect the OARs and the healthy tissues. The amount of influence of each beamlet with unit intensity on the voxels is determined by the dose influence matrix, and the dose delivered to the voxels is obtained by multiplying it by the intensity vector of the beamlets. Therefore, to be able to determine the optimal intensity of beamlets and also downsize the problem, and at the same time to ensure the existence of sample voxels related to all structures in the model, in this article sample voxels with different influence ranges from the dominant beamlet have been selected from the structures which are under the related beamlet influence. Accordingly, for each beamlet bj and in each of the organs at risk under the influence of the bj, the set of voxels for which beamlet bj is dominant, is sorted in ascending order based on the amount of influence received from beamlet bj. Two voxels have been selected from the set, which are the voxels of the second quartile and the fourth quartile, the median voxel, and the voxel with the maximum influence of the bj, respectively. Since dose control in planning target volume (PTV) is more momentous to achieve therapeutic objectives, more voxels of PTV are selected for each of the beamlets. Hence, the first and third quartile voxels have been added to the previous two voxels for PTV. Therefore, two voxels are selected for each beamlet in each of the OARs, and four voxels are selected from the PTV structures.

Algorithm 1 depicts the pseudo-code of the proposed down-sampling algorithm. Down-sampling has been done in two ways in order to compare the results of our proposed method with other research results. In the first way, titled by SMP-1, the voxels have been sampled similar to the previous research by Fountain et al. [14], one voxel corresponding to each beamlet has been selected from each structure. In the second way, called SMP-2 as explained in this section, two voxels for organs and four voxels for tumors are selected, corresponding to each beamlet. Methods SMP-1 and SMP-2 were highlighted with orange and blue in algorithm 1, respectively. The default mode of the pseudo-code in algorithm 1 is based on the SMP-2 sampling, which could simply change to the SMP-1 by replacing the blue highlighted lines with the orange highlighted line.

---

**Algorithm 1. Pseudo Code Of The Proposed Down-Sampling Algorithm**

---

**Requires**: Dose influence matrix; Structures and their related voxels
**Sort** rows of the Dose influence matrix in descending order (sort the matrix horizontally)
**Save** Dominant Beamlet of Voxels: [Max Influence of Voxel, Related beamlet: Dominant Beamlet of Voxel]

**for** all Beamlets
  **for** all Structures
    **for** all Voxels of the current Structure
      **if** the current Beamlet is the Dominant Beamlet of the Voxel
        **save** VSB matrix (Voxels of Structure with current Beamlet as Dominant Beamlet)
      **end**
    **end**
    **if** VSB is not empty
      **sort** Voxels of VSB based on their influence by the Beamlet in ascending order
      **add** fourth quartile and second quartile Voxels as *Sample Voxels* regarding the current Structure and Beamlet
      **if** the current Structure is PTV
        **add** third quartile and first quartile Voxels as *Sample Voxels* regarding the current Structure and Beamlet
      **end**        %SMP-2%
      **clear** VSB
    **end**

    **end**
  **end**
  **unique** the *Sample Voxels*
  **save** the *Sample Voxels*

>     **add** forth quartile Voxels as *Sample Voxels* regarding the current Structure and Beamlet   *%SMP-1%*

### B. Mathematical model

In this paper, a mathematical model is applied that achieved an optimal and clinically applicable treatment plan by using dose prediction and improving it as much as possible toward the prescribed dose. In other words, the model pursues two goals: 1- Achieving the predicted dose and 2- Maximum improvement of the treatment plan toward the prescribed dose. In general, quadratic and linear terms in the objective function have been used to achieve the mentioned goals, respectively. The proposed objective function consists of four parts, each of which contains several terms. The four sections are 1- Optimizing the dose of target volume voxels, 2- Controlling the dose of voxels of OARs, 3- Optimizing the maximum dose of voxels of related OARs, and 4- Optimizing the mean dose of related OARs. More details about the mathematical model have been explained by Yousefi et al. (2023) and the total model is proposed as the following [38]:

$$\min Z = \frac{\psi_1 \sum_{v \in PTV} \omega_v * UD_v^2}{\sum_{v \in PTV} \omega_v} + \frac{\psi_2 \sum_{v \in PTV} \omega_v * OD_v^2}{\sum_{v \in PTV} \omega_v} + \frac{\xi_1 \sum_{v \in PTV} \omega_v |Pres_v - \sum_{b \in B} A_{v,b} X_b|}{\sum_{v \in PTV} \omega_v}$$

$$+ \frac{\psi_3 \sum_{v \in OAR} \omega_v * OD_v^2}{\sum_{v \in OAR} \omega_v} + \frac{\xi_2 \sum_{v \in OAR}(\omega_v \sum_{b \in B} A_{v,b} X_b)}{\sum_{v \in OAR} \omega_v} + \frac{\psi_4 \sum_{v \in OAR\_Max} \omega_v * OD\_MaxP_v}{\sum_{v \in OAR\_Max} \omega_v}$$

$$- \frac{\xi_3 \sum_{v \in OAR\_Max} \omega_v * UD\_MaxP_v}{\sum_{v \in OAR\_Max} \omega_v} + \psi_5 \sum_{S \in OAR\_Mean} OD\_MeanP_S^2 + \xi_4 \sum_{S \in OAR\_Mean} Mean_S^2$$

$S.t$:

$$\sum_{b \in B} A_{v,b} X_b \geq \min\{Pred_v, Pres_v\} - UD_v \qquad \forall\, v \in PTV$$

$$\sum_{b \in B} A_{v,b} X_b \leq \max\{Pred_v, Pres_v\} + OD_v \qquad \forall\, v \in PTV$$

$$\sum_{b \in B} A_{v,b} X_b \leq Pred_v + OD_v \qquad \forall\, v \in OAR$$

$$\sum_{b \in B} A_{v,b} X_b \leq MaxP_S + OD\_MaxP_v \qquad \forall\, v \in S,\ S \in OAR\_Max$$

$$\left( \sum_{b \in B} A_{v,b} X_b - \zeta_S * MaxP_S \right)^+ \leq \chi_S * MaxP_S - UD\_MaxP_v \qquad \forall\, v \in S, S \in OAR\_Max$$

$$Mean_S = \frac{\sum_{v \in S} \sum_{b \in B} \omega_v (A_{v,b} X_b)}{\sum_{v \in S} \omega_v} \qquad \forall\, S \in OAR\_Mean$$

$$Mean_S \leq MeanP_S + OD\_MeanP_S \qquad \forall\, S \in OAR\_Mean$$

$$SPG_{\beta r} \leq 65 \qquad \forall\, \beta \in Beam, r \in Beam\_row$$

$$OD_v, UD_v, OD\_MaxP_v, UD\_MaxP_v \geq 0 \qquad \forall\, v \in \{OAR, PTV\}$$

$$OD\_MeanP_S \geq 0 \qquad \forall\, S \in OAR\_Mean$$

$$X_b \geq 0 \qquad \forall\, b \in B$$

Where $\psi_i$ and $\xi_j$ are importance coefficients in the objective function, and $\omega_v$ is the weight of each voxel $v$. $UD_v$ and $OD_v$ are under-dose and overdose of the voxel $v$, respectively. $Pred_v$ and $Pres_v$ are the predicted dose and the prescribed dose of the voxel $v$, respectively. $B$ is a set of all beamlets $b$, each of which has an intensity of $X_b$. The dose influence matrix is denoted by $A$ and the dose delivered to voxel $v$ by the beamlet $b$ with a unit intensity is shown by $A_{v,b}$. Here, *OAR_Max* and *OAR_Mean* denote the series and parallel OARs with the limitations of the max dose and mean dose of structure, respectively. $OD\_MaxP_v$ and $UD\_MaxP_v$ are overdose and under-dose of the voxel $v$ from the maximum predicted dose of the related structure ($MaxP_S$), respectively. parameters $\zeta_S$ and $\chi_S$ are in the range of (0,1) which $\zeta_S$ is closer to one, and $\chi_S$ is closer to zero and $\chi_S \geq 1 - \zeta_S$, and $OD\_MeanP_S$ is the overdose of the structure $S$ from the mean predicted dose of that structure ($MeanP_S$). The mean dose of organ $S$ is denoted by $Mean_S$ and at last, $SPG_{\beta r}$ is the sum of positive gradient of row $r$ from beam $\beta$.

Finally, values of the importance coefficients of the objective function's terms were allocated similarly to those shown in Table 1 (M: Million, k: kilo). The suggested down-sampling algorithm as well as the mathematical model are applied to real data and the results are presented and discussed in the following sections.

**Table 1. values of the objective function's importance coefficients**

| Coefficient | $\psi_1$ | $\psi_2$ | $\psi_3$ | $\psi_4$ | $\psi_5$ | $\xi_1$ | $\xi_2$ | $\xi_3$ | $\xi_4$ |
|---|---|---|---|---|---|---|---|---|---|
| Value | 2M | 0.5M | 0.2M | 0.2M | 1k | 20k | 0.2k | 1k | 50 |

### C. Data and analysis program

In this paper, the open-access knowledge-based planning dataset (OpenKBP) [39] has been applied. The OpenKBP provides an augmented variation of real clinical data. Specifically, patient data is provided from several institutions that is available on The Cancer Imaging Archive (TCIA) [40], which hosts several open-source datasets. The current dataset was established in order to use in the OpenKBP grand challenge (February to June 2020) and completed with the 21 series of results (i.e., dose predictions) of the challenge participants. Data for 340 patients were provided who were treated for head-and-neck cancer with 6 MV step-and-shoot IMRT in 35 fractions to satisfy the prescribed dose to the high-risk target (i.e., PTV70) from nine equispaced coplanar beams at angles 0, 40, 80, . . ., 320 degrees. The data is split into training (n=200), validation (n=40), and testing (n=100) sets. Every patient in these datasets has a dose distribution, CT images, structure masks, a feasible dose mask (i.e., the mask of where the dose can be non-zero), and voxel dimensions. Moreover, 21 series of dose predictions for each patient of testing sets (n=100) have been provided by the challenge participants which 19 series have better prediction quality (19 different predictions for each of the 100 patients). Therefore, complete real data for 100 patients along with 21 series of dose predictions have been shared publically in the OpenKBP dataset. Dose influence matrices were generated using the same parameters in the IMRTP library from A Computational Environment for Radiotherapy Research (CERR) [41].

In this paper, we used the same 30 randomly selected sets in Yousefi et al. (2023) from the 100 testing sets (from set 241 to 340) of the Open KBP dataset, so the results of two researches can be properly compared. Hence, real data of 30 patients with head and neck cancer and their 19 sets of better-quality dose prediction, constitute the data of the current paper. The random patients selected via MATLAB software are sets: 245, 249, 251, 253, 254, 259, 262, 265, 270, 282, 292, 293, 294, 296, 299, 300, 301, 303, 307, 308, 309, 312, 316, 317, 319, 329, 330, 334, 335 and 339.

To solve the suggested quadratic-linear mathematical model (*QuadLin*) we used CVX, a Matlab-based convex modeling framework for specifying and solving convex programs [42, 43]. An academic CVX professional license has been applied to use CVX with commercial solver Mosek. Matlab software of version R2018a was applied in this paper on a single computer with an Intel Core i5-8250U (1.80 GHz) CPU and 16 GB of RAM.

### D. Plan evaluation approaches

In the current paper, models are evaluated and compared regarding two main fields which are the quality of the treatment plan as well as the computation efficiency. In order to measure the quality and performance of generated treatment plans in comparison to reference plan, two evaluation approaches are used in this research. One of them is DVH point differences, which is the absolute difference between a DVH point from dose distributions and the reference plan. The DVH differences are evaluated on two and three DVH points for each organ-at-risk (OAR) and planning target volume (PTV), respectively. The OAR DVH points were the $D_{mean}$ and $D_{0.1cc}$, which was the mean dose delivered to the OAR and the maximum dose delivered to 0.1cc of the OAR, respectively. The PTV DVH points were the $D_{1\%}$, $D_{95\%}$, and $D_{99\%}$, which were the dose delivered to 1% (99$^{th}$ percentile), 95% (5$^{th}$ percentile), and 99% (1$^{st}$ percentile) of voxels in the PTV, respectively. The second evaluation approach regarding measuring plan quality is clinical criteria satisfaction, which is defined as the proportion of clinical criteria that were satisfied by a treatment plan. This approach focuses on one criterion of each region of interest (ROI) as can be seen in table 2. Furthermore, the computational efficiency of the models is compared based on their solve time.

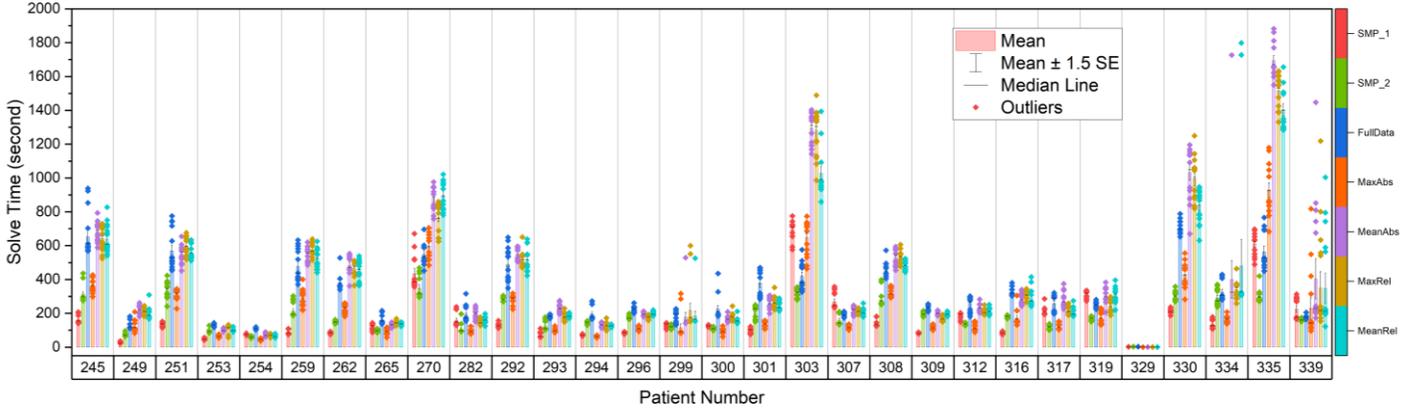

**Fig. 1. Box plot of seven modes solve times for each of the random 30 patients**

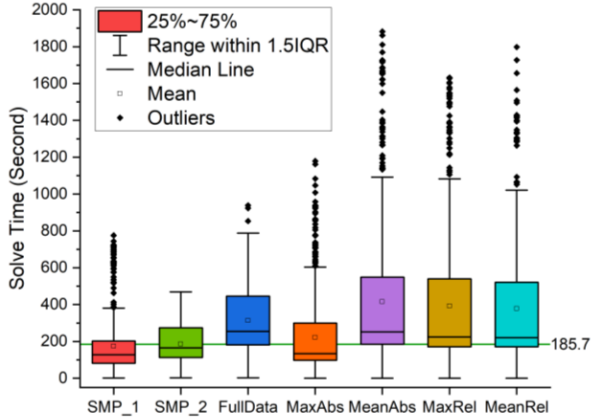

**Fig. 2. Box chart of seven modes solve times**

| Table 2. clinical criteria for each ROI | |
|---|---|
| ROIs | Criteria |
| OARs | |
| Brainstem | $D_{0.1cc} \leq 50.0$ Gy |
| Spinal cord | $D_{0.1cc} \leq 45.0$ Gy |
| Right parotid | $D_{mean} \leq 26.0$ Gy |
| Left parotid | $D_{mean} \leq 26.0$ Gy |
| Esophagus | $D_{mean} \leq 45.0$ Gy |
| Larynx | $D_{mean} \leq 45.0$ Gy |
| Mandible | $D_{0.1cc} \leq 73.5$ Gy |
| PTVs | |
| PTV56 | $D_{99\%} \geq 53.2$ Gy |
| PTV63 | $D_{99\%} \geq 59.9$ Gy |
| PTV70 | $D_{99\%} \geq 66.5$ Gy |

### III.   Results

Sampling method similar to Fountain et al. [14] (*SMP-1*) and our proposed down-sampling technique (*SMP-2*) were implemented on the data of 30 random patients from the Open KBP dataset. After that, the *QuadLin* optimization model was solved for three modes of data including *Full-Data*, *SMP-1*, and *SMP-2* for each of 30 patients with 19 dose predictions for each patient, consequently, 1710 problems were solved. Next, the resulting plans of the above three modes, along with the outputs of four models presented by Babier et al. [37], including MaxAbs, MeanAbs, MaxRel, and MeanRel for the same 30 patients, are evaluated regarding the computation efficiency (i.e. solve time) as well as the quality of plans.

#### A.   Solve time

In this section, seven solutions are compared regarding their solving time. The solutions include two down-sampled data *SMP-1* and *SMP-2*, as well as *Full-Data*, all three of which are optimized using the QuadLin model; in addition to outputs of the MaxAbs, MeanAbs, MaxRel, and MeanRel models. Box plot of seven modes solve times has been illustrated in Figure 1, where its horizontal axis shows the number of random 30 patients. Figure 2 has displayed an overall comparison of the seven solutions regarding the solve time mean and median by box chart. Based on figures 1 and 2, the proposed solution SMP-2 is much more stable in solve times with no outliers and a minimum spread. Also, the mean and median of solve time for SMP-2 are 185.7 and 165.4 seconds, respectively. Accordingly, SMP-2 has located in second place regarding the solve time, after the SMP-1 with a mean solve time of 174 seconds. In order to have a comprehensive comparison of the computing efficiency, it is necessary to compare the solutions from the perspective of treatment quality including DVH points and clinical criteria satisfaction, in addition to the solution time.

### B. DVH points

To compare the relative quality of dose distributions and review the performance of the proposed down-sampling method, DVH point differences between the reference plan and three modes of QuadLin-produced dose, including *Full-Data*, *SMP-2*, and *SMP-1* were calculated. The differences were evaluated over the five DVH points listed in section II-D. Accordingly, SMP-2 is better than SMP-1 for all DVH points of the PTVs, i.e. $D_{95\%}$, $D_{99\%}$, and $D_{1\%}$, and half DVH points of the OARs, i.e. $D_{mean}$ and $D_{0.1cc}$. Figure 3 illustrates a sample DVH diagram for a patient and a prediction set comparing three modes of QuadLin-generated plans including Full-Data, SMP-2, and SMP-1 with the predicted dose. Based on figure 3, SMP-2 has a performance quite close to Full-Data and is much better than SMP-1 and the prediction.

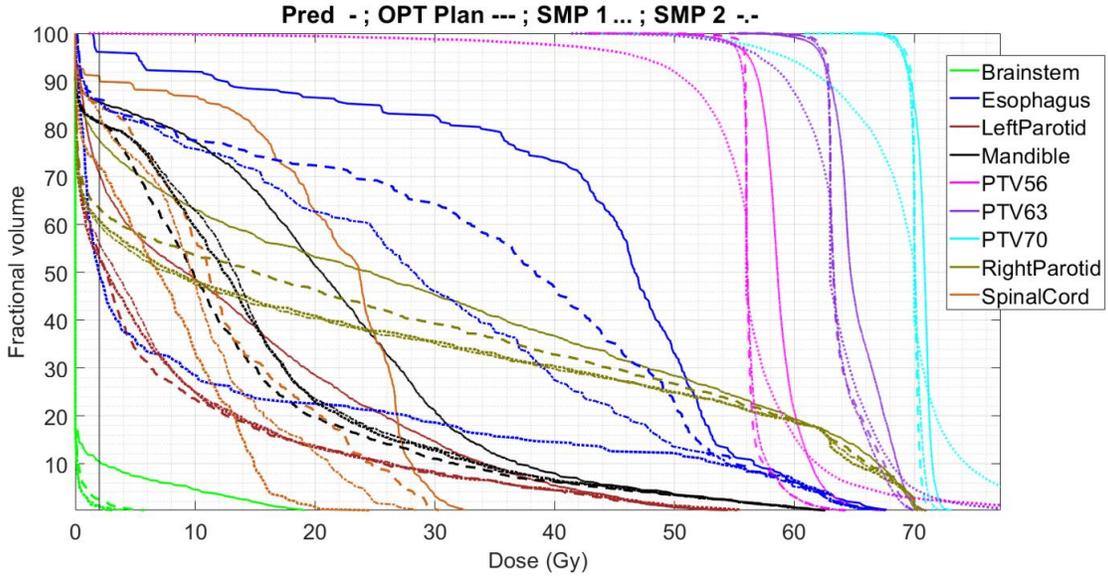

**Fig. 3. Sample DVH diagram for a patient and a prediction set comparing three modes of *QuadLin*-generated plans including Full-Data (OPT Plan), SMP-2, SMP-1 with the predicted dose (Pred)**

### C. Clinical criteria satisfaction

In this approach, the proportion of the clinical criteria that were satisfied by three QuadLin-generated plans, i.e. *Full-Data*, *SMP-2*, and *SMP-1*, as well as the predicted plans for random 30 patients are calculated separately for each of the 19 prediction sets. Clinical criteria consist of one criterion of each ROI as shown in table 2, which are $D_{mean}$ for parallel OARs, $D_{0.1cc}$ for series OARs, and $D_{99\%}$ for the targets. Figure 4 visualized the percentage of all ROIs satisfied criteria for the predicted plans (orange bullets) and the plans generated by the QuadLin model using three modes of data including complete data (Full-Data, blue bullets) and two modes of down-sampling data (SMP-1 and SMP-2, red and green bullets respectively); with a horizontal dashed line represents the reference plan situation. The results indicate that on average for 570 problems (30 patients with 19 prediction sets), the proposed *SVSIDB* down-sampling method, i.e. *SMP-2* mode with green bullets, improved the percentage of clinical criteria satisfaction by more than 18% and 22.5% compared to the predicted dose (orange bullets) and the SMP-1 sampling method (red bullets) which presented previously by Fountain et al. [14], respectively; while it diluted the Full-Data results (blue bullets) only by 3%. Hence, *SMP-2* has performed far better than *SMP-1* regarding clinical criteria satisfaction.

## IV. Discussion

Optimizing the IMRT treatment plan is a time-consuming problem because of the large number of voxels. In order to reduce the size of the problem, most studies presented random down-sampling methods and some researchers provided several kinds of voxel clustering techniques. In the current paper, a novel method is proposed for down-sampling the large-scale problems of IMRT treatment planning, titled *SVSIDB*. The results of the proposed method titled SMP-2 is compared to the results of three recent researches, including a sampling method by Fountain et al. [14] called SMP-1, Yousefi et al. [38] called Full-Data, and the four models MaxAbs, MeanAbs, MeanAbs, and MaxRel from Babier et al. [37]. Figure 5 illustrates the percentage of satisfied clinical criteria for all ROIs by the above seven models. As expected, the Full-Data has the best results for all 19 prediction sets. SMP-2 is at a little distance after that with an average of more than 81%, and it is interesting to note that it performed better than the reference line for all sets. This means that the poor quality of dose prediction (i.e. orange bullets) specially in case of sets 17, 18, and 19 has not been able to have a great impact on the performance of the SMP-2 and did not cause it to fall below the reference line of 69%. The four models presented by Babier et al. (2022) have an average performance on both sides of the reference line and depend on the prediction quality. Fountain et al.'s SMP-1 model has poor performance and is more than 10% lower than the reference line. Table 3 shows the solve time and clinical criteria of the SMP-2 in comparison to the mentioned six models. The mean solve time of the SMP-2 is 185.7 seconds which is 46% lower than the Full-Data and on average 43% lower than the four models of Babier et al. [37]. Although the average solve time of SMP-1 is only 7% better than SMP-2, its satisfied clinical criteria are by far lower than SMP-2 (more than 22%), which seriously questions the computational efficiency and performance of SMP-1. Since Fountain et al. [14] established and evaluated their model based on a small number of patients with a simple problem that consisted of just one OAR, one tumor volume, and two beams, the applicability of their presented method in case of complicated problems is doubtful. According to table 3, the percentage of satisfied clinical criteria of SMP-2 is 81.3% which is 22.5% higher than SMP-1 and on average 12% better than Babier et al.'s four models. The SMP-2 has not only reduced the solve time but also improved the quality of the treatment plans. This is a remarkable achievement of the proposed model compared to the previous researches by Fountain et al. and Babier et al. [14, 37]. Furthermore, proposed down-sampling algorithm has the potential of better performance in the field of treatment quality by re-weighting the objective function of the mathematical model. The subject of automatic weight adjustment of applied QuadLin mathematical model is being studied as a field of future research.

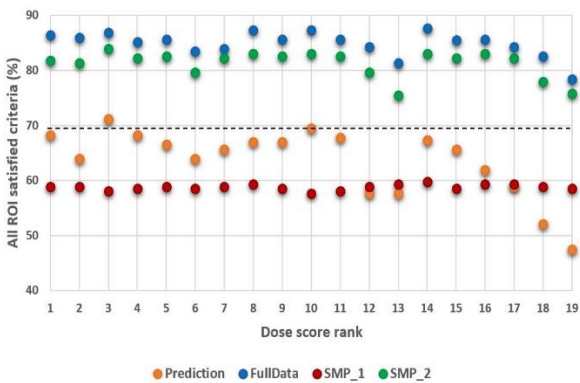

**Fig. 4. Percentage of satisfied clinical criteria for all ROIs by 3D$^3$ prediction (orange bullets) compared with three data modes QuadLin- generated plans including SMP-1 (red bullets), SMP-2 (green bullets), and Full-Data (blue bullets). The dashed line depicts the reference plan situation.**

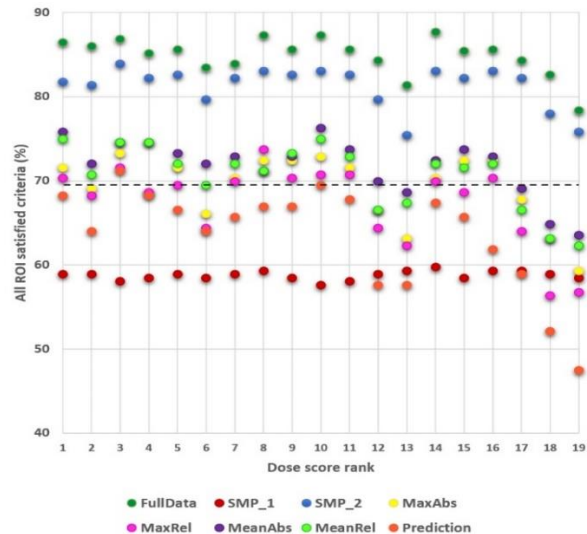

**Fig. 5. Percentage of satisfied clinical criteria for all ROIs by all models**

**Table 3. Comparing the solve time and clinical criteria of the SMP-2 with the other six models**

| Model | Average solve time (second) | Percentage of reducing solve time by SMP-2 | Percentage of satisfied clinical criteria (all ROIs) | satisfied clinical criteria (all ROIs) improving by SMP-2 |
|---|---|---|---|---|
| SMP-2 (Proposed method) | 185.7 | --- | 81.3% | --- |
| SMP-1 (Fountain et al. [14]) | 174 | -7% | 58.8% | 22.5% |
| Full-Data (Yousefi et al. [38]) | 341 | 46% | 84.9% | -3.6% |
| MaxAbs (Babier et al. [37]) | 222 | 16% | 69.2% | 12.1% |
| MeanAbs (Babier et al. [37]) | 389 | 52% | 71.8% | 9.5% |
| MaxRel (Babier et al. [37]) | 393 | 53% | 67.4% | 13.9% |
| MeanRel (Babier et al. [37]) | 367 | 49% | 70.7% | 10.6% |

## IV. Conclusion

The current paper contributed to knowledge-based planning and improved the computational efficiency of IMRT treatment planning. The proposed down-sampling method titled SVSIDB with SMP-2 mode considerably reduced the solve time and significantly improved the clinical criteria. The results of the current study indicated a considerable improvement in the computation efficiency compared to the results of previous research while at the same time outperforming the plan evaluation indicators. Accordingly, the proposed method reduced the average solving time by 46% in comparison to the full-data QuadLin model. The results also show an up to 53% reduction in solve time along with up to 22% improvement in clinical criteria compared to the previous research. This is a remarkable achievement of the proposed model compared to the previous research and confirmed the effectiveness of the *SVSIDB* method which has the potential of even more improvement of the computation efficiency.

## Acknowledgments
The first author would like to present this article to his kind wife and intelligent daughter.

## Conflict of Interest Statement
The authors have no relevant conflicts of interest to disclose.

## References

[1] Mahnam, M., Gendreau, M., Lahrichi, N., & Rousseau, L.M. (2017). Simultaneous delivery time and aperture shape optimization for the volumetric-modulated arc therapy (VMAT) treatment planning problem. Physics in Medicine and Biology, 62(14), 5589–5611.

[2] Kufer, K.H., Scherrer, A., Monz, M., & Thieke, C. (2003). Intensity modulated radiotherapy— a large scale multi-criteria programming problem. OR Spectrum, 25, 223–249.

[3] Scherrer, A., Kufer, K.H., Bortfeld, T., Monz, M., & Alonso, F. (2005). IMRT planning on adaptive volume structures—A decisive reduction in computational complexity. Physics in Medicine and Biology, 50(9), 2033.

[4] Scherrer, A., & Kufer, K. H. (2008). Accelerated IMRT plan optimization using the adaptive clustering method. Linear Algebra and Its Applications, Special Issue on Linear and Nonlinear Models and Algorithms in Intensity-Modulated Radiation Therapy, 428 (5–6), 1250–1271, doi:10.1016/j.laa.2007.03.025.

[5] Ferris MC, Einarsson R, Jiang Z, Shepard D. 2006. Sampling issues for optimization in radiotherapy. Ann Oper Res. 148(1):95–115.

[6] Martin, B. C., Bortfeld, T. R., & Castañon, D. A. (2007). Accelerating IMRT optimization by voxel sampling. Physics in Medicine and Biology, 52(24), 7211.

[7] Zarepisheh, M., Ungun, B., Ruijiang, L., Yinyu, Y., Boyd, S., & Lei, X. (2018). Advances in Inverse Planning Algorithm and Strategy, Chapter 6 of Advanced and Emerging Technologies in Radiation Oncology Physics, Kim S, Wong J, CRC Press, ISBN 9781498720045, 9780429508141.

[8] Lan Y, Li C, Ren H, Zhang Y, Min Z. 2012. Fluence map optimization (FMO) with dose-volume constraints in IMRT using the geometric distance sorting method. Phys Med Biol. 57(20):6407–6428.

[9] Potrebko PS, Fiege J, Biagioli M, Poleszczuk J. 2017. Investigating multi-objective fluence and beam orientation IMRT optimization. Phys Med Biol. 62(13):5228–5244.

[10] Yousefi A, Ahmadi S. (2022). "Improving the Computational Efficiency of Cancer Treatment Planning Optimization", Proceeding of 7[th] International Conference on Combinatorics, Cryptography, Computer science and Computing (I4C), IUST, Tehran, Iran.

[11] Yousefi A. (2023). "Optimizing Electrical Energy Consumption in Medical Sector: Improving the Computational Efficiency and Reducing Treatment Time in Radiotherapy for Cancer", Proceeding of 5[th] International Conference on Optimizing Electrical Energy Consumption (OEEC), Feb. 28-Mar. 1, Amirkabir University of Technology, Tehran, Iran.

[12] Yousefi, A, Hadi-Vencheh, A, (2023). Resiliency and reliability of the power grid in the time of COVID-19: An integrated ABC-K-means model for optimal positioning of repair crew, Electric Power Systems Research, Vol. 216, 109022, https://doi.org/10.1016/j.epsr.2022.109022.

[13] Yang Y, Xing L. 2018. Autonomous treatment plan optimization strategy augmented by using a knowledge-guided and irregularly downsampled voxelization scheme. Int J Radiat Oncol Biol Phys. 102(3):S236.

[14] Fountain, L., Khedriliraviasl, K., Mahmoudzadeh, S., & Mahmoudzadeh, H. (2021). Dose-based constraint generation for large-scale IMRT optimization. INFOR: Information Systems and Operational Research, DOI: 10.1080/03155986.2021.2004636.

[15] Momin S, Fu Y, Lei Y, et al. Knowledge-based radiation treatment planning: A data-driven method survey. J Appl Clin Med Phys. 2021;22:16–44. https://doi.org/10.1002/acm2.13337

[16] Zarepisheh M, Long T, Li N, et al., 2014. A DVH-guided IMRT optimization algorithm for automatic treatment planning and adaptive radiotherapy replanning, Medical Physics 41(6), 061711 [PubMed: 24877806].

[17] Chanyavanich V, Das SK, Lee WR, Lo JY. Knowledge-based IMRT treatment planning for prostate cancer. Med Phys. 2011;38:2515-2522.

[18] Good D, Lo J, Lee WR, Wu QJ, Yin F-F, Das SK. A knowledge-based approach to improving and homogenizing intensity modulated radiation therapy planning quality among treatment centers: an example application to prostate cancer planning. Int J Radiat Oncol Biol Phys. 2013;87:176-181.

[19] Nwankwo O, Sihono DSK, Schneider F, Wenz F. A global quality assurance system for personalized radiation therapy treatment planning for the prostate (or other sites). Phys Med Biol. 2014;59:5575.

[20] Schmidt M, Lo JY, Grzetic S, Lutzky C, Brizel DM, Das SK. Semiautomated head-and-neck IMRT planning using dose warping and scaling to robustly adapt plans in a knowledge database containing potentially suboptimal plans. Med Phys. 2015;42:4428-4434.

[21] Wang B, Lei Y, Tian S, et al. Deeply supervised 3D fully convolutional networks with group dilated convolution for automatic MRI prostate segmentation. Med Phys. 2019;46:1707-1718.

[22] Harms J, Lei Y, Wang T, et al. Paired cycle-GAN-based image correction for quantitative cone-beam computed tomography. Med Phys. 2019;46:3998-4009.

[23] Fu Y, Mazur TR, Wu X, et al. A novel MRI segmentation method using CNN-based correction network for MRI-guided adaptive radiotherapy. Med Phys. 2018;45:5129-5137.

[24] Dong X, Lei Y, Wang T, et al. Automatic multiorgan segmentation in thorax CT images using U-net-GAN. Med Phys. 2019;46:2157-2168.

[25] Yousefi, A., Ketabi, S., Abedi, I. (2022). How to Apply $3D^3$ Prediction? A Novel Mathematical Model to Generate Pareto Optimal Clinical Applicable IMRT Treatment Plan On the Foundation of Dose Prediction and Prescription. arXiv preprint, arXiv:2206.05834 [math.OC]. https://doi.org/10.48550/arXiv.2206.058

[26] Vasant K, Chan JW, Tianqi W, et al. DoseGAN: a generative adversarial network for synthetic dose prediction using attention-gated discrimination and generation. Sci Rep. 2020;10.

[27] Nguyen D, Long T, Jia X, et al. A feasibility study for predicting optimal radiation therapy dose distributions of prostate cancer patients from patient anatomy using deep learning. Sci Rep. 2017;9:1-10.

[28] Ge Y, Wu QJ. Knowledge-based planning for intensity-modulated radiation therapy: a review of data-driven approaches. Med Phys. 2019;46:2760-2775.

[29] McIntosh C, Welch M, McNiven A, Jaffray DA, Purdie TG. Fully automated treatment planning for head and neck radiotherapy using a voxel-based dose prediction and dose mimicking method. Phys Med Biol. 2017;62:5926.

[30] Fan Jiawei, Jiazhou Wang, Zhi Chen, Chaosu Hu, Zhen Zhang, and Weigang Hu. (2019) "Automatic treatment planning based on three-dimensional dose distribution predicted from deep learning technique", Med. Phys. 46 (1).

[31] Babier A., R. Mahmood, A. L. McNiven, A. Diamant, and T. C. Y. Chan. (2020) "The importance of evaluating the complete automated knowledge-based planning pipeline," Phys Med, vol. 72, pp. 73–79.

[32] Zhong Y, Yu L, Zhao J, Fang Y, Yang Y, Wu Z, Wang J and Hu W (2021) Clinical Implementation of Automated Treatment Planning for Rectum Intensity-Modulated Radiotherapy Using Voxel-Based Dose Prediction and Post-Optimization Strategies. Front. Oncol. 11:697995. doi: 10.3389/fonc.2021.697995

[33] Nilsson V, Gruselius H, Zhang T, De Kerf G and Claessens M (2021). Probabilistic dose prediction using mixture density networks for automated radiation therapy treatment planning. Phys. Med. Biol. 66.5, p. 055003.

[34] Eriksson O, Zhang T (2022). Robust automated radiation therapy treatment planning using scenario-specific dose prediction and robust dose mimicking. Med. Phys. 49.6, pp. 3564-3573.

[35] Zhang T, Bokrantz B, Olsson J (2021). Probabilistic feature extraction, dose statistic prediction and dose mimicking for automated radiation therapy treatment planning. Med. Phys. 48.9, pp. 4730-4742.

[36] Zhang T, Bokrantz B, Olsson J (2022). Probabilistic Pareto plan generation for semiautomated multicriteria radiation therapy treatment planning. Phys. Med. Biol. 66, p. 045001.

[37] Babier A, Mahmood R, Zhang B, Alves VGL, Barragán-Montero AM, Beaudry J, Cardenas CE, et al. OpenKBP-Opt: an international and reproducible evaluation of 76 knowledge-based planning pipelines. Phys Med Biol. 2022 Sep 12;67(18). DOI: 10.1088/1361-6560/ac8044. PMID: 36093921.

[38] Yousefi A, Ketabi S, Abedi I. A novel mathematical model to generate semi-automated optimal IMRT treatment plan based on predicted 3D dose distribution and prescribed dose. Med Phys. 2023;1-11. https://doi.org/10.1002/mp.16236

[39] Babier A., B. Zhang, R. Mahmood, K. L. Moore, T. G. Purdie, A. L. McNiven, and T. C. Y. Chan, "OpenKBP: The open-access knowledge-based planning grand challenge and dataset", Med Phys, vol. 48, no. 9, pp. 5549–5561, 2021.

[40] Clark K., Bruce Vendt, Kirk Smith, John Freymann, Justin Kirby, Paul Koppel, Stephen Moore, Stanley Phillips, David Matt, Michael Pringle, Lawrence Tarbox, and Fred Prior. (2013) "The cancer imaging archive (TCIA): maintaining and operating a public information repository". J Digit Imaging, 26(6):1045-1057. doi: 10.1007/s10278-013-9622-7.

[41] Deasy J. O., A. I. Blanco, and V. H. Clark. CERR: a computational environment for radiotherapy research. Med Phys, 30(5): 979-85, 2003.

[42] Michael Grant and Stephen Boyd. CVX: Matlab software for disciplined convex programming, version 2.0 beta. http://cvxr.com/cvx, September 2013.

[43] Michael Grant and Stephen Boyd. Graph implementations for nonsmooth convex programs, Recent Advances in Learning and Control (a tribute to M. Vidyasagar), V. Blondel, S. Boyd, and H. Kimura, editors, pages 95-110, Lecture Notes in Control and Information Sciences, Springer, 2008. http://stanford.edu/~boyd/graph_dcp.html.